\documentclass{aastex}
\usepackage{spr-astr-addons}
\usepackage{url}
\RequirePackage{color}

\begin{document}

\title{Lowering the Self-Coupling of the Scalar Field in the Generalized Higgs Inflation}
\slugcomment{Not to appear in Nonlearned J., 45.}
\shorttitle{Short article title} \shortauthors{Nozari et al.}

\author{Kourosh Nozari\altaffilmark{1}}\quad \affil{Department of Physics, Faculty of Basic Sciences,
University of Mazandaran, P. O. Box 47416-95447, Babolsar, IRAN}
\affil{Research Institute for Astronomy and Astrophysics of Maragha
(RIAAM), P. O. Box 55134-441, Maragha, Iran}\\
\author{Somayeh Shafizadeh\altaffilmark{2}}\affil{Department of Physics,
Payame Noor University (PNU), P. O. Box 19395-3697, Tehran, Iran}
\and
\author{Narges Rashidi\altaffilmark{3}} \affil{Department of Physics, Faculty of Basic Sciences,
University of Mazandaran, P. O. Box 47416-95447, Babolsar, IRAN}
\and
\altaffiltext{1}{knozari@umz.ac.ir}
\altaffiltext{2}{s.shafizadeh@tpnu.ac.ir}
\altaffiltext{3}{n.rashidi@umz.ac.ir}

\begin{abstract}
We study cosmological dynamics of a generalized Higgs inflation. By
expanding the action up to the second and third order in the small
perturbations, we study the primordial perturbation and its
non-Gaussian distribution. We study the non-Gaussian feature in both
the equilateral and orthogonal configurations. By adopting a quartic
potential, we perform a numerical analysis on the model's parameter
space and compare the results with Planck2015 observational data. To
obtain some observational constraint, we focus on the self-coupling
and the non-minimal coupling parameters. We show that, in the
presence of the non-minimal coupling and the Galileon-like
interaction, the self-coupling parameter can be reduced to the
order of $10^{-6}$ which is much larger than the value that CMB normalization suggests for this self-coupling.
\end{abstract}

\section{Introduction}

Cosmological inflation is a part of the cosmic history related to a
homogeneous and isotropic FRW universe that expands almost
exponentially (a nearly de-Sitter universe) at very early stage of
the universe evolution. The simplest model of inflation is the one
in which a single scalar field with the almost flat potential runs
the cosmic inflation. The theory of inflation is successful to
address some problems of the standard cosmological model as well as
to provide the initial density perturbations seeding the large scale
structures~(Guth 1981, Linde 1982, Albrecht and Steinhardt 1982,
Linde 1990, Liddle and Lyth 2000, Lidsey et al. 1997, Rioto 2000,
Lyth and Liddle 2009). The nearly scale invariant, adiabatic and
Gaussian distribution of the perturbations modes is one of the
notable predictions of the simple single field inflation (Maldacena
2003). However, by proposing some extended models of inflation and
considering the non-linear perturbations, it is possible to predict
some level of non-Gaussianity of the primordial
perturbations~(Maldacena 2003, Bartolo et al. 2004, Babich et al.
2004, Seery and Lidsey 2005, Cheung et al. 2008, Chen 2010, De
Felice and Tsujikawa 2011a, De Felice and Tsujikawa 2011b, De Felice
et al. 2011, Nozari and Rashidi 2012, Nozari and Rashidi 2013a,
Nozari and Rashidi 2013b, Nozari and Rashidi 2014, Nozari and
Rashidi 2016a, Nozari and Rashidi 2016b, Nozari and Rashidi 2017).
So, it is reasonable to expect a level non-Gaussianity in future
observation.

The discovery of Higgs as a fundamental particle in electro-weak
symmetry breaking has significant implication in particle physics
and cosmology. Regarding this fact that most of the inflation models
require a scalar field (inflaton) to explain the accelerating
expansion of the early universe, there is a possibility that Higgs
field to be a good candidate for the inflaton~(See for instance
Barbon and Espinosa 2009, Calmet et al. 2017 and references
therein). However, when the Higgs field is minimally coupled to the
gravity, its self-coupling is too large to achieve the slow-roll
inflation. Actually, to suppress the amplitude of the curvature
perturbation (which should be much smaller than the Planck scale),
we need to reduce the self-coupling of the Higgs field. To this end,
some extensions of the Higgs inflation model have been proposed. One
of these extensions is the model in which the Higgs field is
non-minimally coupled to the gravity sector of the theory. In this
model, a large amount of the non-minimal coupling parameter
effectively suppresses the self-coupling of the Higgs field
(Futamase and Maeda 1989, Salopek et al, 1989, Fakir and Unruh 1990,
Kaiser 1995, Tsujikawa and Gumjudpai 2004, Bezrukov et al. 2009,
Bezrukov and Shposhnikov 2008, Bezrukov and Shposhnikov 2009,
Barvinsky et al. 2009, Watanabe 2011). Unfortunately, it seems that
the non-minimal model violates the unitarity bound (Burgess et al.
2009, Lerner and McDonald 2010, Germani and Kehagias 2010).

It should be mentioned that the loop corrections are all
small compared to the tree level amplitude and since the
inflationary energy scale is always much below the scale of
unitarity violation (to get constraint $\lambda\ll1$), it is not
needed to worry about the stability of inflationary model in this
context (see Germani and Kehagias 2010, Calmet and Casadio 2014 for
more details).

Another extension of the Higgs inflation is the new Higgs inflation
model which relies on the non-minimal derivative coupling between
the scalar field and Einstein tensor (Amendola 1993). By this
coupling, since the normalization of the inflation field is changed,
the magnitude of the Higgs self-coupling could be lower than its
experimental bound.

Also in this case, due to presence of the non-renormalizable
operator in the new Higgs inflationary action the time dependence
unitarity bound is set. By requiring the scale of curvature that is
much lower than the unitarity bound, we impose Hubble parameter
scale below the Planck scale. Therefore, this postulated coupling is
free of unitarity bound during inflation (Barbon and Espinosa 2009,
Germani and Kehagias 2010, Atkins and Calmet 2011).

Another approach is the running kinetic inflation model in which the
non-canonical kinetic term changes the normalization of the Higgs
field and smoothes the general steep potential (Takahashi 2010, Dimopoulos and Thomas 2003).

The phenomenological features of the running kinetic inflation have
been studied in (Nakayama and Takahashi 2008) with details. Higgs-G
inflation also, is an extension of the standard Higgs inflation
which incorporates the higher order derivatives of the scalar field
(Kobayashi et al. 2010, Kamada et al. 2011). In the Galileon model
the Lagrangian is formulated in a such way that the field equations
are invariant under the Galileon symmetry
$\partial_{\mu}\varphi\rightarrow\partial_{\mu}\varphi+a_{\mu}$ in
the Minkowski limit of the theory (Nicolis et al. 2009, De Felice
and Tsujikawa 2012). Note that, the expression $X\Box\varphi$ which
is introduced as the Galileon term, emerges in the DGP model as a
consequence of the combination of a brane-bending mode and a
transverse graviton (Deffayet et al. 2002, Porrati 2002, Luty et al.
2003). By adding the Galileon term to the theory, the potential
essentially becomes flat and the quantum fluctuations are
suppressed.

In this paper, we consider another class of the generalized Galileon
Higgs inflation which is a subclass of the most generalized
scalar-tensor theory (Deffayet et al. 2011, Charmousis et al. 2012).
Considering that the chaotic inflation is not confirmed properly by
the observational data (Komatsu et al. 2010, Amsler et al. 2008), it
is interesting to adopt the quartic potential and make the theory
observationally viable (Germani and Kehagias 2010).

By considering an inflation model with the Galileon effect, enhanced
kinetic term and non-minimal coupling between the Higgs filed and
both the scalar and tensor parts of the gravity, we try to reduce
the self-coupling of the Higgs sector. In doing so, we preserve also the
cosmological viability of the setup. Actually the energy scale of the Higgs
self-coupling constant, $\lambda$, is in the interval
$0.11<\lambda<0.27$. From the CMB normalization, $\lambda$ is
constrained to be of the order of $10^{-13}$~(Liddle and Lyth 2000)-
the scale that Higgs boson can't reach. However, by considering an
inflation model with the Galileon effect, enhanced kinetic term and
non-minimal coupling between the Higgs field and both the scalar and
tensor parts of the gravity, we try to reduce the energy scale (self-coupling) of
the Higgs sector. As we shall see, by considering this extended model, we are able to
reduce the energy scale of the Higgs self-coupling constant,
$\lambda$, from interval $0.11<\lambda<0.27$ to less than $10^{-6}$.
So, in this paper, our aim is to reduce $\lambda$ by
considering the Galileon-like and non-minimal effects and keeping
the observational viability of the model's parameters. In this
regard, by decreasing the order of $\lambda$ and approaching the
energy scale of the inflation era, the Higgs field can be considered
to be an inflaton. We note that the negative values of $\lambda$ are
possible in essence and at least theoretically. However, in this case
inflation never happens which is out of our interest in this paper.

With these explanations, in section 2, we introduce the generalized
Higgs G-inflation model and the action of the theory. In section 3
we study the background dynamics of the model. In section 4, by
adopting the ADM formalism, we expand the action up to the second
and third orders of the perturbations. In this section we obtain the
scalar and tensor spectral index of the primordial perturbations. We
also study the non-Gaussian feature of the perturbations in both
equilateral and orthogonal configurations. After that, in section 5
we perform a numerical analysis on the model's parameter space and
compare the results with Planck2015 data set. In this regard, we
obtain some constraints on the model's parameters.

\section{Generalized Higgs G-Inflation Model}

By detecting the Higgs boson in Large Hadron collider (LHC)
experiment in Geneva (Chatrchyan et al. 2012, Aad et al. 2012), many efforts have been
made to construct the inflation models where the Higgs boson acts as
an inflaton. In this respect, the following Lagrangian is devoted to
Higgs boson in the absence of gravity
\begin{eqnarray}
\label{eq1}{\cal{L}}_{\cal{H}}=-{\cal{D}}_{\mu}{\cal{H}}^{\dagger}{\cal{D}}^{\mu}{\cal{H}}-
\lambda\big({\cal{H}}^{\dagger}{\cal{H}}-\upsilon^{2}\big)^{2}\,,
\end{eqnarray}
where ${\cal{D}}_{\mu}$ is the covariant derivative corresponding to
the SM gauge symmetry. ${\cal{H}}$ is the Higgs boson and
$\upsilon\sim 246$ GeV is its expectation value. Since the parameter
$\upsilon$ is very small compared with the Higgs field during
inflation era, we can safely eliminate this parameter. Also, we
concentrate on the radial part of the Higgs boson,
$\phi\sim\sqrt{2{\cal{H}}^{\dagger}{\cal{H}}}$, and ignore the
contributions of the gauge sectors of the SM~(Germani et al. 2014). In this
regard, in the presence of the gravity, the Lagrangian of the Higgs
model takes the following form
\begin{equation}
\label{eq2}{\cal{L}}=\frac{m^{2}_{pl}}{2}R-\frac{1}{2}\partial_{\mu}\varphi\partial^{\mu}\varphi-
\frac{\lambda}{4}\varphi^{4}\,.
\end{equation}
Unfortunately, due to the large value of the Higgs self-coupling,
this model in not viable (Bezrukov and Shaposhnikov 2008).
Therefore, it seems reasonable to focus on the generalized
G-inflation models. The action of the generalized G-inflation is
written as
\begin{equation}
\label{eq3}S=\sum^{4}_{i=2}\int d^{4}x\sqrt{-g}{\cal{L}}_{i}\,,
\end{equation}
where $g$ is the determinant of the metric $g_{\mu\nu}$ and
\begin{equation}
\label{eq4}{\cal{L}}_{2}=K(\varphi,X)\,,
\end{equation}
\begin{equation}
\label{eq5}{\cal{L}}_{3}=-G_{3}(\varphi,X)\Box\varphi\,,
\end{equation}
\begin{equation}
\label{eq6}{\cal{L}}_{4}=G_{4}(\varphi,X)R+G_{4,X}\Bigg[(\Box\varphi)^{2}-(\nabla_{\mu}\nabla_{\nu}\varphi)^{2}\Bigg]\,.
\end{equation}
In the above equations, $R=6(\dot{H}+2H^{2})$ is the Ricci scalar,
$K$ and $G_{i}$ are arbitrary functions of $\varphi$ and
$X=-(\frac{1}{2})g^{\mu\nu}\nabla_{\mu}\varphi\nabla_{\nu}\varphi$.
We define
$G_{i}(\varphi,X)=g_{i}(\varphi)+h_{i}(\varphi)X$. In fact, in this
definition we have expanded $G_{i}(\varphi,X)$ as
$G_{i}(\varphi,X)=g_{i}(\varphi)+h_{i}(\varphi)X+k_{i}(\varphi)X^2+l_{i}(\varphi)X^{3}+...$
and just kept the terms up to the first order in $X$ and ignored the
higher order ones. We also have $\Box\equiv
g^{\mu\nu}\nabla_{\mu}\nabla_{\nu}$ (the standard d'Alembertian
operator). This theory has been originally found by Horndeski in a
different form~(Horndeski1974). This generalized model consists of
the running kinetic term, the Galileon interaction, the non-minimal
coupling and the non-minimal derivative coupling of the scalar field
and gravity. We adopt the arbitrary functions $K(\varphi,X)$ and
$G_{i}(\varphi,X)$ as\footnote{Note that, in the running kinetic inflation model,
rapid growth of the kinetic term at large values of inflaton field
causes the potential to be flat. In fact, in paper (Nakayama and
Takahashi 2008) it has been discussed that the coefficient of the
kinetic term is not necessarily unity. Actually, when the inflaton
rolls over a large scale in high-scale inflation model, this
coefficient is not close to 1. In this regard, to cover this issue,
it is appropriate to consider the general kinetic term in the action
of the model.}
\begin{equation}
\label{eq7}K(\varphi,X)={\cal{K}}(\varphi)X-V\,,
\end{equation}
\begin{equation}
\label{eq8}G_{3}(\varphi,X)=\gamma(\varphi)X\,,
\end{equation}
\begin{equation}
\label{eq9}G_{4}(\varphi,X)=\frac{1}{2}(m^{2}_{pl}+\xi\varphi^{2})+\frac{1}{2\mu^{2}}X\,,
\end{equation}
where $\xi$ is a dimensionless non-minimal coupling parameter,
$\gamma(\varphi)$ is a dimensionless function of the Higgs field and
$\mu$ is a mass scale. The function $G_{3}$ in equation (8) has been
chosen in the way that we cover the coupling between the scalar
field, kinetic term and the second-order derivatives of the
scalar field (Galileon gravity). Also, in equation (9), the minimal
and non-minimal coupling of the gravity with the scalar field and
the coupling between the gravity and derivatives of the field have
been considered. Note that, if we set ${\cal{K}}=1$,
$G_{3}(\varphi,X)=0$ and $G_{4}(\varphi,X)=\frac{m_{pl}^{2}}{2}$,
then the standard Higgs inflation is recovered.

\section{The Background Dynamics}

To derive the background equations of the model, we consider the FRW
background specified by the metric
$ds^{2}=-dt^{2}+a^{2}(t)\delta_{ij}dx^{i}dx^{j}$. By varying action
$(3)$ with respect to the metric, we find the following Friedmann
equations
\begin{equation}
\label{eq10}H^{2}=\frac{\rho_{\varphi}}{3m^{2}_{pl}}\,,\,\,\,\,\,\
\dot{H}=-\frac{\rho_{\varphi}+p_{\varphi}}{2m^{2}_{pl}} \,,
\end{equation} where the energy density and the pressure of the
scalar field are defined as
\begin{eqnarray}
\label{eq11}\rho_{\varphi}=\frac{1}{2}\dot{\varphi}^{2}\Bigg[{\cal{K}}+6\gamma\
\frac{H\dot{\varphi}}{m^{2}_{pl}}+\gamma_{,\varphi}\frac{\dot{\varphi}^{2}}{m^{2}_{pl}}+\frac{9}{\mu^{2}}
\frac{H^{2}}{m^{2}_{pl}}\nonumber\\+12\xi\frac{H\varphi}{\dot{\varphi}}+6\xi\frac{H^{2}\varphi^{2}}{\dot{\varphi}^{2}}\Bigg]
+V \,,
\end{eqnarray}
\begin{eqnarray}
\label{eq12}p_{\varphi}=\frac{1}{2}\dot{\varphi}^{2}\Bigg[{\cal{K}}-\gamma_{,\varphi}
\frac{\dot{\varphi}^{2}}{m^{2}_{pl}}-2\gamma
\frac{\ddot{\varphi}}{m^{2}_{pl}}-\frac{4}{m^{2}_{pl}}(\frac{H\ddot{\varphi}}
{\mu^{2}\dot{\varphi}})\nonumber\\-\frac{1}{m^{2}_{pl}}\bigg(\frac{3H^{2}+2\dot{H}}{\mu^{2}}\bigg)+4\xi
+2\big(\frac{\xi\varphi^{2}}{\dot{\varphi}^{2}}-1\big)
(3H^{2}+2\dot{H})
\nonumber\\+4m^{2}_{pl}\xi\varphi\Big(\frac{\ddot{\varphi}}{\dot{\varphi}^{2}}+\frac{2H}{\dot{\varphi}}\Big)\Bigg]-V\,,
\end{eqnarray}
respectively. The equation of motion, obtained by varying the action
$(3)$ with respect to $\varphi(t)$, is given by
\begin{equation}
\label{eq13}\frac{1}{a^{3}}\frac{d}{dt}(a^{3}J)=P_{\varphi}\,,
\end{equation}
where
\begin{equation}
\label{eq14}J\equiv\dot{\varphi}({\cal{K}}+\frac{3}{\mu^{2}}H^{2})+3\gamma
H\dot{\varphi}^{2}-\gamma_{,\varphi}\dot{\varphi}^{3}\,,
\end{equation}
and
\begin{equation}
\label{eq15}P_{\varphi}=-V'-\frac{1}{2}\dot{\varphi}^{2}\bigg(2\gamma_{,\varphi}\ddot{\varphi}+
\gamma_{,\varphi\varphi}\dot{\varphi}^{2}-2\frac{\xi\varphi}{\dot{\varphi}^{2}}R
\bigg)\,.
\end{equation}
By substituting equations $(14)$ and $(15)$ into the equation $(13)$
we get
\begin{eqnarray}
\label{eq16}\ddot{\varphi}\Big({\cal{K}}+6\gamma
H\dot{\varphi}+\frac{3}{\mu^{2}}H^{2}-2\gamma_{,\varphi}\dot{\varphi}^{2}\Big)
+3H\dot{\varphi}\Big({\cal{K}}+3\gamma
H\dot{\varphi}\nonumber\\+\frac{3}{\mu^{2}}H^{2}\Big)
+\frac{1}{2}\dot{\varphi}\Big({\cal{K}}_{,\varphi}+6\gamma\dot{H}-\gamma_{,\varphi\varphi}\dot{\varphi^{2}}\Big)
\nonumber\\-6\xi\varphi(2H^{2}+\dot{H})+V'=0\,.\hspace{1cm}
\end{eqnarray}

The slow-roll conditions in this setup are as follows
\begin{eqnarray}
\label{eq18}\dot{\varphi}^{2}\ll V(\varphi),\,\
\mid\ddot{\varphi}\mid \ll \mid H\dot{\varphi}\mid,
\mid\dot{{\cal{K}}}\mid \ll \mid H{\cal{K}}\mid,\,\,\nonumber\\ \mid
\dot{g}_{i}(\varphi)\mid \ll \mid Hg_{i}(\varphi)\mid,\,\ \mid
\dot{h}_{i}\mid \ll Hh_{i}(\varphi)\,.
\end{eqnarray}
By considering the slow-roll conditions, the Ricci scalar becomes
\begin{equation}
\label{eq19}R\simeq\frac{1}{\big[{m^{2}_{pl}}+\xi(1+6\xi)\varphi^{2}\big]}\big[4V(\varphi)+6\xi\varphi
V' \big]\,.
\end{equation}
Also, the main background equations within the slow-roll limits take
the following form
\begin{equation}
\label{eq20}H^{2}\simeq \frac{V}{3(m^{2}_{pl}+\xi\varphi^{2})}\,,
\end{equation}
\begin{equation}
\label{eq21}3H\dot{\varphi}\bigg ({\cal{K}}+3 H
\dot{\varphi}\gamma+\frac{3}{\mu^{2}}H^{2}\bigg)-\xi R
\varphi\simeq-V'\,.
\end{equation}
From equation (20) we see that in the generalized G-inflation the
friction term is enhanced. By using equations $(19)$ and $(20)$ we
can derive $\frac{d\varphi}{d N}\equiv\frac{\dot{\varphi}}{H}$ as
follows
\begin{eqnarray}
\label{eq22}\frac{d\varphi}{d N}=\frac{\dot{\varphi}}{H} \simeq
\hspace{5cm}
\nonumber\\-\frac{2V'_{eff}}{Y\bigg({\cal{K}}+(\frac{1}{\mu^{2}})Y+\sqrt{({\cal{K}}+(\frac{1}{\mu^{2}})Y)^{2}
-4\gamma V'_{eff}\bigg)}}\,,
\end{eqnarray}
where
\begin{eqnarray}
\label{eq23} V'_{eff}\equiv\hspace{7cm} \nonumber\\
\frac{1}{(m_{pl}^{2}+\xi(1+6\xi)\varphi^{2})}\big[-4\xi\varphi
V(\varphi)+(m_{pl}^{2}+\xi\varphi^{2})V'\big]\,,
\end{eqnarray}
\begin{equation}
Y\equiv\frac{V}{(m^{2}_{pl}+\xi\varphi^{2})}\,,
\end{equation}
and $N=\ln a$ is the number of e-folds parameter. If the expression
in the bracket of equation (22) to be small compared with the
denominator, we have slow-roll inflation even with a steep
potential.

The slow-roll parameters in our setup are obtained as follows
\begin{eqnarray}
\label{eq24}\alpha=\frac{\dot{g}_{4}(\varphi)}{Hg_{4}(\varphi)}\simeq V_{eff}' \times\hspace{4cm}\nonumber\\
\Bigg(\frac{-4\xi\varphi V^{-1}}{{\cal{K}}+
(\frac{Y}{\mu^{2}})+\sqrt{({\cal{K}}+(\frac{Y}{\mu^{2}}))^{2}-4\gamma
V_{eff}'}}\Bigg);\,\,\,\,\alpha\ll 1
\end{eqnarray}
\begin{eqnarray}
\label{eq25}\epsilon=-\frac{\dot{H}}{H^{2}}\simeq V^{-1}Y^{-1}\times\hspace{3.5cm}\nonumber\\
\Bigg(\frac{V'^{2}_{eff}}
{{\cal{K}}+\frac{Y}{\mu^{2}}+\sqrt{({\cal{K}}+(\frac{Y}{\mu^{2}}))^{2}-4\gamma
V_{eff}'}}\Bigg)-\frac{\alpha}{2};\,\,\,\,\epsilon\ll 1
\end{eqnarray}
\begin{equation}
\label{eq26}\eta\simeq\epsilon-\frac{1}{2\epsilon}(\frac{d\epsilon}{dN});\,\,\,\,\eta\ll
1
\end{equation}
where
\begin{eqnarray}
\label{eq27}\frac{d\epsilon}{dN}=\hspace{6cm}\nonumber\\
\Bigg(\frac{-2Y^{-1}V'_{eff}}{{\cal{K}}+
(\frac{1}{\mu^{2}})Y+\sqrt{({\cal{K}}+(\frac{1}{\mu^{2}})Y)^{2}
-4\gamma V'_{eff}}}\Bigg)\frac{d\epsilon}{d\varphi}\,,
\end{eqnarray}
and
\begin{eqnarray}
\label{eq28}\zeta=\frac{\dot{J}}{HJ}\simeq\hspace{6cm}\nonumber\\
\epsilon-\Bigg(\frac{V''_{eff} Y^{-1}}{{\cal{K}}+
\frac{Y}{\mu^{2}}+\sqrt{({\cal{K}}+\frac{Y}{\mu^{2}})^{2} -4\gamma
V'_{eff}}}\Bigg);\,\,\,\,\zeta\ll 1\,.
\end{eqnarray}
The number of e-folds during inflation which is given by
\begin{equation}
\label{eq29}N=\int Hdt=\int\frac{H}{\dot{\varphi}}d\varphi\,,
\end{equation}
in the generalized G-inflation model takes the following form
\begin{eqnarray}
\label{eq30}N\simeq\hspace{6cm}\nonumber\\
\int\frac{-2\gamma
Y}{\bigg({\cal{K}}+\frac{Y}{\mu^{2}}+\sqrt{\big({\cal{K}}+\frac{Y}{\mu^{2}}\big)^{2}
-4\gamma V'_{eff}}\bigg)}d\varphi\,.
\end{eqnarray}

\section{Perturbation and Non-Gaussianity}

In this section, we study the perturbations in our setup. To study
the tensor and scalar parts of the perturbations we should expand
the action up to the second order. We work in the unitary gauge
($\delta\varphi=0$) and adopt the ADM formalism with the following
metric (Baumann 2009, Mukhanov 1992)
\begin{equation}
\label{eq31}ds^{2}=-N^{2}dt^{2}+\gamma_{ij}(dx^{i}+N^{i}dt)(dx^{j}+N^{j}dt)\,,
\end{equation}
where $N$ and $N^{i}$ are the lapse and shift functions. In metric
(31) we have the following definition
\begin{eqnarray}
\label{eq32}N=1+2\Phi,\,\,N_{i}=\delta_{ij}\partial^{j}B,\,\,\nonumber\\
\gamma_{ij}=a^{2}(t)(1+2\Psi)(\delta_{ij}+h_{ij})\,.
\end{eqnarray}
$\Phi, \Psi,$ and $B$ are the scalar perturbations and $h_{ij}$ is
the spatial shear 3-tensor. Now, we rewrite the perturbed metric up
to the linear level as (Baumann 2009, Mukhanov 1992)
\begin{eqnarray}
\label{eq33}ds^{2}=-(1+2\Phi)dt^{2}+2a^{2}(t)B_{,i}dx^{i}dt\nonumber\\+a^{2}(t)(1+2\Psi)(\delta_{ij}+h_{ij})dx^{i}dx^{j}\,.
\end{eqnarray}
By considering the scalar part of this metric, we expand the action
$(3)$ up to second order in the small perturbations and get
\begin{eqnarray}
\label{eq34}S_{2}=\int
dtd^{3}xa^{3}\Bigg\{-3\Big((m^{2}_{pl}+\xi\phi^{2})-\frac{X}{2\mu^{2}}\Big)\dot{\Psi}^{2}
\nonumber\\+\frac{1}{a^{2}}\bigg[2\Big((m^{2}_{pl}+\xi\phi^{2})-\frac{X}{2\mu^{2}}\Big)\dot{\Psi}-
\bigg(2H(m^{2}_{pl}+\xi\phi^{2})-\nonumber\\ \frac{6}{\mu^{2}}H
X+2\dot{\phi}(\xi\phi-\gamma
X\big)\bigg)\Phi\bigg]\partial^{2}B-\frac{2}{a^{2}}\Big((m^{2}_{pl}+\xi\phi^{2})\nonumber\\-\frac{X}{2\mu^{2}}\Big)
\Phi\partial^{2}\Psi+3\bigg(2H(m^{2}_{pl}+\xi\phi^{2})-\frac{6}{\mu^{2}}H
X\nonumber\\+2\dot{\phi}(\xi\phi-\gamma
X\big)\bigg)\Phi\dot{\Psi}+\bigg[X\Big({\cal{K}}+12\gamma\dot{\phi}H+
\frac{18}{\mu^{2}}\Big)\nonumber\\-4\gamma_{\varphi}X^{2}-3H^{2}(m^{2}_{pl}+\xi\phi^{2})-
6H\xi\phi\dot{\phi}\bigg]\Phi^{2}+\nonumber\\
\frac{1}{a^{2}}\Big((m^{2}_{pl}+\xi\phi^{2})+\frac{X}{\mu^{2}}\Big)(\partial\Psi)^{2}\Bigg\}\,.
\end{eqnarray}
By using the above second order action, we can find the momentum and
Hamiltonian constrains as
\begin{equation}
\label{eq35}\Phi=L_{1}\dot{\Psi}\,,
\end{equation}
where
\begin{equation}
\label{eq36}L_{1}=\frac{2m^{2}_{pl}+\xi\phi^{2}-\frac{X}{\mu^{2}}}{H\Bigg(2m^{2}_{pl}+\xi\varphi^{2}-
7\frac{X}{2\mu^{2}}\Bigg)-\gamma
X\dot{\varphi}+m^{2}_{pl}\xi\varphi\dot{\varphi}}\,,
\end{equation}
and
\begin{eqnarray}
\label{eq37}\frac{1}{a^{2}}\partial^{2}B=3\dot{\Psi}-\frac{1}{a^{2}}L_{1}\partial^{2}\Psi+\hspace{2.6cm}\nonumber\\
\frac{X\Big({\cal{K}}+12\gamma\dot{\phi}H+
\frac{18}{\mu^{2}}\Big)-4\gamma_{\phi}X^{2}} {-\frac{3}{\mu^{2}}H
X+H(2m^{2}_{pl}+\xi\phi^{2})+\dot{\phi}(m^{2}_{pl}\xi\phi-\gamma
X\big)}\Phi \nonumber\\
-\frac{3H^{2}(2m^{2}_{pl}+\xi\phi^{2})-12m^{2}_{pl}H\xi\phi\dot{\phi}}
{-\frac{3}{\mu^{2}}H
X+H(2m^{2}_{pl}+\xi\phi^{2})+\dot{\phi}(m^{2}_{pl}\xi\phi-\gamma
X\big)}\Phi\,.
\end{eqnarray}
By substituting the equation $(35)$ in equation $(34)$ and
integrating it by parts, the second order action reduces to the
following expression
\begin{equation}
\label{eq38}S_{2}=\int
dtd^{3}xa^{3}{\cal{U}}\Big[\dot{\Psi}^{2}-\frac{c_{s}^{2}}{a^{2}}(\partial\Psi)^{2}\Big]\,,
\end{equation}
where
\begin{eqnarray}
\label{eq39}{\cal{U}}\equiv \Bigg[2\Big(m^{2}_{pl}+\xi\phi^{2}
-\frac{X}{\mu^{2}}\Big)^{2}\bigg[X\Big({\cal{K}}+16\gamma\dot{\phi}H\nonumber\\
+
\frac{18}{\mu^{2}}H^{2}+4\gamma_{\phi}X\Big)-3H^{2}(2m^{2}_{pl}+\xi\phi^{2})\nonumber\\-6m^{2}_{pl}H\xi\phi\dot{\phi}\bigg]\Bigg]
\Bigg(\Big[-6H\frac{X}{\mu^{2}}+2H(2m^{2}_{pl}+\xi\phi^{2}
\nonumber\\+\frac{X}{\mu^{2}})+\dot{\phi}(-2\gamma
X+2m^{2}_{pl}\xi\phi)\Big]^{2}\Bigg)^{-1}\nonumber\\+3\Bigg(2m^{2}_{pl}+\xi\phi^{2}-\frac{X}{\mu^{2}}\Bigg)\,,\hspace{0.1cm}
\end{eqnarray}
and
\begin{equation}
\label{eq40}c^{2}_{s}\equiv\frac{4-\frac{2({\cal{K}}+(\frac{1}{\mu^{2}})Y)}
{\bigg({\cal{K}}+(\frac{1}{\mu^{2}})Y+\sqrt{({\cal{K}}+(\frac{1}{\mu^{2}})Y)^{2}
-4\gamma
V'_{eff}}\bigg)}}{3\Bigg[2-\frac{2({\cal{K}}+(\frac{1}{\mu^{2}})Y)}
{\bigg({\cal{K}}+(\frac{1}{\mu^{2}})Y+\sqrt{({\cal{K}}+(\frac{1}{\mu^{2}})Y)^{2}
-4\gamma V'_{eff}}\bigg)}\Bigg]}\,.
\end{equation}
To avoid the ghosts and gradient instabilities it is required that
\begin{equation}
\label{eq41}{\cal{U}}>0,\,\,\,\,\,\,c^{2}_{s}>0.
\end{equation}
Actually, there are two constraints on the sound speed of
the perturbations (Ellis 2007, Quiros 2017): 1- The squared sound
speed of the perturbations ($c_{i}^{2}$ with $i=s,T$ where $T$
denotes the tensor part of the perturbations which we'll study
later) should be positive in order to avoid the appearance of
Laplacian instabilities. That is, $c_{i}^{2}>0$. 2: From the
causality requirement, the sound speed of the perturbations should
be smaller than (at most, equal to) the local speed of light. That
means, $c_{i}^{2}\leq c^{2}$. Since in this paper we set
$c=1$, the constraint becomes $c_{i}^{2}\leq 1$. These constraints
are satisfied in equations (41), as we will see in the numerical
analysis of non-Gaussianities. The constraints on $c_{s}^2$, lead to
positive equilateral configuration of non-Gaussianities and negative
orthogonal configuration of non-Gaussianities.

For convenience we define the following parameters
\begin{equation}
\label{eq42}F=2+\frac{1}{m^{2}_{pl}}\Big(\xi\varphi^{2}+\frac{X}{\mu^{2}}\Big)\,,
\end{equation}
\begin{equation}
\label{eq43}\epsilon_{s}=\epsilon+\frac{\dot{\varphi}}{m^{2}_{pl}HF}\Big(\gamma
X+\xi\varphi\Big)\,.
\end{equation}
By these definitions, we have the relation
$\epsilon_{s}=\frac{{\cal{U}} c_{s}^{2}}{m_{pl}^{2}F}$. Causality,
Laplacian and ghost free requirements impose that the right hand
side of this equation to be positive. Therefore the left hand side
of the relation should be positive too. On the other hand,
$\dot{H}>0$ means $\epsilon <0$. In the case of negative $\epsilon$,
to have positive $\epsilon_{s}$, the second term of equation (43)
should be large enough (actually larger than $|\epsilon|$). Since
this model is an extended one, it is likely possible to find some
parameter space that gives $\dot{H}>0$. However, in this paper we
don't look after this case.

The power spectrum of the curvature perturbations is given by
\begin{equation}
\label{eq44}{\cal{P}}_{\Psi}=\frac{H^{2}}{8\pi^{2}{\cal{U}}c^{3}_{s}}\,.
\end{equation}
With this definition, we obtain the scalar spectral index as
\begin{eqnarray}
\label{eq45}n_{s}-1=\frac{d\ln{\cal{P}}_{\Psi}}{d\ln
k}\Bigg|_{c_{s}k=aH } =-4\epsilon+\eta-\zeta+\nonumber\\
\frac{V'_{eff}\bigg(\frac{{\cal{K}}+(\frac{1}{\mu^{2}})Y}
{{\cal{K}}+(\frac{1}{\mu^{2}})Y+\sqrt{({\cal{K}}+(\frac{1}{\mu^{2}})Y)^{2}
-4\gamma V'_{eff}}}\bigg)'} {Y\bigg({\cal{K}}+
(\frac{1}{\mu^{2}})Y+\sqrt{({\cal{K}}+(\frac{1}{\mu^{2}})Y)^{2}
-4\gamma V'_{eff}}\bigg)} \nonumber\\
\times\Biggr[\frac{1}{2-\frac{2({\cal{K}}+(\frac{1}{{\mu^{2}}})Y)}
{\bigg({\cal{K}}+(\frac{1}{\mu^{2}})Y+\sqrt{({\cal{K}}+(\frac{1}{\mu^{2}})Y)^{2}
-4\gamma V'_{eff}}\bigg)}}-\nonumber\\
\frac{3}{4-\frac{2({\cal{K}}+(\frac{1}{{\mu^{2}}})Y)}
{\bigg({\cal{K}}+(\frac{1}{\mu^{2}})Y+\sqrt{({\cal{K}}+(\frac{1}{\mu^{2}})Y)^{2}
-4\gamma V'_{eff}}\bigg)}}\Biggr]\,.
\end{eqnarray}
Now, we consider the tensor part of the metric (33) and expand the
quadratic action for the tensor perturbations as follows
\begin{equation}
\label{eq46}S_{T}=\int
dtd^{3}xa^{3}\big[{\cal{Q}}\dot{h}^{2}_{ij}-\frac{\cal{F}}{a^{2}}(\partial
h_{ij})^{2}\big]\,,
\end{equation}
where
\begin{equation}
\label{eq47}{\cal{Q}}\equiv(m^{2}_{pl}+\xi\varphi^{2})-\frac{X}{\mu^{2}}\,,\,\,
{\cal{F}}\equiv2(m^{2}_{pl}+\xi\varphi^{2})+\frac{X}{\mu^{2}}\,.
\end{equation}
The sound speed square is given by
\begin{equation}
\label{eq48}c^{2}_{T}\equiv\frac{\cal{F}}{\cal{Q}}\equiv\frac{m^{2}_{pl}+\xi\varphi^{2}+\frac{X}{\mu^{2}}}
{m^{2}_{pl}+\xi\varphi^{2}-\frac{X}{\mu^{2}}}\,.
\end{equation}
Note that, satisfying conditions ${\cal{Q}}>0$ and $c_{T}^{2}>0$
lead to the ghost and Laplacian free perturbations. We note that constraint from
observation of gravitational waves by LIGO/VIRGO opens a research area at this point.

The power spectrum of primordial tensor perturbations is given by
\begin{equation}
\label{eq49}{\cal{P}}_{T}=\frac{H^{2}}{2\pi^{2}{\cal{Q}}c^{3}_{T}}\simeq
\Bigm(\frac{1}{2m^{2}_{pl}+2\xi\varphi^{2}+\frac{X}{\mu^{2}}}\Bigm)\frac{2H^{2}}{\pi^{2}}\,,
\end{equation}
leading to the following tensor spectral index
\begin{eqnarray}
\label{eq50}n_{T}=\frac{d \ln {\cal{P}}_{T}}{d \ln k}
\simeq\bigg(\frac{V'_{eff}}{V}\bigg)^{2}\times \hspace{2cm} \nonumber\\
\Bigg(\frac{m^{2}_{pl}+\xi\varphi^{2}}{{\cal{K}}+
(\frac{1}{\mu^{2}})Y+\sqrt{({\cal{K}}+(\frac{1}{\mu^{2}})Y)^{2}
-4\gamma V'_{eff}}}\Bigg)\,.
\end{eqnarray}
The tensor-to-scalar ratio in this setup is given by
\begin{eqnarray}
\label{eq51}r=\frac{{\cal{P}}_{T}}{{\cal{P}}_{\Psi}}\simeq
\hspace{5cm} \nonumber\\ -\frac{8}{3\sqrt{3}}
\frac{\Bigg(4-\frac{2({\cal{K}}+(\frac{1}{{\mu^{2}}})Y)}
{{\cal{K}}+(\frac{1}{\mu^{2}})Y+\sqrt{({\cal{K}}+(\frac{1}{\mu^{2}})Y)^{2}
-4\gamma
V'_{eff}}}\Bigg)^{\frac{3}{2}}}{\Bigg(2-\frac{2({\cal{K}}+(\frac{1}{{\mu^{2}}})Y)}
{{\cal{K}}+(\frac{1}{\mu^{2}})Y+\sqrt{({\cal{K}}+(\frac{1}{\mu^{2}})Y)^{2}
-4\gamma V'_{eff}}}\Bigg)^{\frac{1}{2}}}\,n_{T}\,.
\end{eqnarray}
As we see, the consistency relation is modified in the presence of
Galileon effect. If we set $G_{3}=0$, $G_{4}=\frac{m_{pl}}{2}$  and
${\cal{K}}(\varphi)=1$ in equation (51), the model recovers the
standard consistency relation $r=-8n_{T}$.

By regarding this fact that for a Gaussian distribution, any odd
point correlation functions vanishes, to seek for the non-Gaussian
feature we should study three point correlation
function~(Ohashi et al. 2013). To this end, we expand action up to
the third order in the small perturbations. We eliminate parameter
$B$ by using equation (35) and introduce $(\chi)$ as
\begin{eqnarray}
B=-L_{1}\Psi+\frac{a^{2}{\cal{X}}}{\big(2m^{2}_{pl}+\xi\varphi^{2}-\frac{X}{\mu^{2}}\big)}\,,\nonumber
\end{eqnarray}
\begin{eqnarray}
{\cal{X}}=\frac{\Big(2(m^{2}_{pl}+\xi\frac{\varphi^{2}}{2})-\frac{X}{\mu^{2}}\Big)B}{a^{2}}\,.\nonumber
\end{eqnarray}
In this regard, we obtain the cubic action as
\begin{equation}
\label{eq52}S_{3}=\int dt{\cal{L}}_{3}\,,
\end{equation}
where

\begin{eqnarray}
\label{eq53}{\cal{L}}_{3}=\int
d^{3}x\Biggr\{a^{3}\frac{F\epsilon_{s}}{c^{2}_{s}}\Biggm(-3\Big(\frac{1}{c^{2}_{s}}-1\Big)+\frac{1}{c^{2}_{s}}
\Big(\epsilon-\frac{\dot{\epsilon}_{s}}{H\epsilon_{s}}\nonumber\\+\frac{\dot{\varphi}}{m^{2}_{pl}HF}\big(\xi\varphi-3\gamma
X\big)
-6\frac{X}{m^{2}_{pl}\mu^{2}F}\Big)\Biggm)m^{2}_{pl}\Psi\dot{\Psi}^{2}
\nonumber\\+a\Biggl(F\epsilon_{s}\big(\frac{1}{c^{2}_{s}}-1\big)
+\frac{F\epsilon_{s}}{c^{2}_{s}}\Biggl(\epsilon_{s}+\frac{\dot{\epsilon}_{s}}{H
\epsilon_{s}}-2\frac{\dot{c}_{s}}{Hc_{s}}+\nonumber\\
\Bigg(\frac{1}{2m^{2}_{pl}+\xi\varphi^{2}+\frac{X}{\mu^2}}\Bigg)\frac{2X}{\mu^{2}}\Biggl)\Biggl)
m^{2}_{pl}\Psi(\partial\Psi)^{2}\nonumber\\
+a^{3}\Biggm(\frac{m_{pl}F\epsilon_{s}}{Hc^{2}_{s}}\Big(\frac{1}{c^{2}_{s}}-1-2\frac{\Lambda}{\Sigma}\Big)
+\frac{1}{c^{2}_{s}}\Bigg(\frac{\gamma
X\dot{\varphi}}{m^{2}_{pl}HF}\nonumber\\+2\frac{X}{m^{2}_{pl}\mu^{2}F}
-\frac{\xi\varphi\dot{\varphi}}{m^{2}_{pl}HF}\Bigg)-3\frac{\gamma
X\dot{\varphi}}{m^{2}_{pl}HF}+\frac{\xi\varphi\dot{\varphi}}{m^{2}_{pl}HF}
\nonumber\\-2\frac{X}{m^{2}_{pl}\mu^{2}F}-6\frac{c^{2}_{s}}{\epsilon_{s}}\Big(\frac{\gamma
X\dot{\varphi}}{m^{2}_{pl}HF}\Big)^{2}\Biggm)m_{pl}\dot{\Psi}^{3}
-2a^{3}\frac{\epsilon_{s}}{c^{2}_{s}}\nonumber\\
\dot{\Psi}(\partial_{i}\Psi)(\partial_{i}{\cal{X}})
+a^{3}\Big(\frac{1}{4Fm^{2}_{pl}}\big(\epsilon_{s}-4\frac{\gamma
X\dot{\varphi}}{m^{2}_{pl}HF}\big)\Big)\nonumber\\
\partial^{2}\Psi(\partial{\cal{X}})^{2}+a\Big(\frac{2\gamma X\dot{\varphi}}{H^{3}}\Big)\dot{\Psi}^{2}\partial^{2}\Psi
+\Big(-\frac{2}{3}\frac{\gamma X\dot{\varphi}}{H^{3}a}\Big)\nonumber\\
\Big[\partial^{2}\Psi(\partial\Psi)^{2}-\Psi\partial_{i}\partial_{j}(\partial_{i}\Psi)\Big]+\nonumber\\
a\Big(2\frac{\gamma
X\dot{\varphi}}{FH^{2}}\Big)\Big(\partial^{2}\Psi\partial_{i}\Psi\partial_{i}{\cal{X}}-
\Psi\partial_{i}\partial_{j}\Big(\partial_{i}{\cal{X}})\Big)
\Biggr\}\,.
\end{eqnarray}
In this equation, parameters $\Sigma$ and $\Lambda$ are defined as
\begin{eqnarray}
\label{eq54}\Sigma\equiv\frac{\Big(m^{2}_{pl}F-2\frac{X}{\mu^{2}}\Big)}{m^{4}_{pl}}\Biggl[3\Big(m^{2}_{pl}HF
-\gamma
X\dot{\varphi}-4\frac{HX}{\mu^{2}}\nonumber\\+m^{2}_{pl}\xi\varphi\dot{\varphi}\Big)^{2}
+\Big(m^{2}_{pl}F-2\frac{X}{\mu^{2}}\Big)
\Bigg(-3m^{2}_{pl}H^{2}F+X\nonumber\\ +12\gamma
HX\dot{\varphi}-4\gamma_{,\varphi}X^{2}+21\frac{H^{2}X}{\mu^{2}}-6m^{2}_{pl}\xi
H\varphi\dot{\varphi}\Bigg) \Biggl]\,,
\end{eqnarray}
and
\begin{equation}
\label{eq55}\Lambda\equiv\Bigg(2+\frac{1}{m^{2}_{pl}}\bigg(\xi\varphi^{2}+\frac{X}{2\mu^{2}}\bigg)\Bigg)^{2}
\Bigr[\gamma
HX\dot{\varphi}-\frac{4}{3}\gamma_{,\varphi}X^{2}\Bigr]\,.
\end{equation}
To obtain the three point correlators, we should calculate the
vacuum expectation value of the curvature perturbations during
inflation as follows (see for instance (Maldacena 2003, Cheung et
al. 2008, Seery and Lidsey 2005))
\begin{eqnarray}
\label{eq56}\langle\Psi(\mathbf{k}_{1})\Psi(\mathbf{k}_{2})\Psi(\mathbf{k}_{3})\rangle=\hspace{4cm}\nonumber\\
-i\int^{\tau_{f}}_{\tau_{i}}d\tau
a\langle0|[\Psi(0,\mathbf{k}_{1})\Psi(0,\mathbf{k}_{2})\Psi(0,\mathbf{k}_{3}),{\cal{H}}_{int}(\tau)]|0\rangle\,,
\end{eqnarray}
where interacting Hamiltonian is ${\cal{H}}_{int}=-{\cal{L}}_{3}$.
We can assume that the dimensionless coefficient of each
contribution in the third order action can be treated as a constant
because of the slow varying of those coefficients during the
inflation epoch. In this respect, by solving the integral (56) we
get
\begin{equation}
\label{eq57}\langle\Psi_{\mathbf{k}_{1}}\Psi_{\mathbf{k}_{2}}\Psi_{\mathbf{k}_{3}}\rangle=
(2\pi)^{3}\delta(\mathbf{k}_{1}+\mathbf{k}_{2}+\mathbf{k}_{3})\emph{\cal{B}}_{\Psi}(k_{1},k_{2},k_{3})\,,
\end{equation}
with
\begin{equation}
\label{eq58}\emph{B}_{\Psi}=\frac{(2\pi)^{4}}{4\prod^{3}_{i=1}k^{3}_{i}}({\cal{P}}_{\Psi})^{2}
{\cal{A}}_{\Psi}(k_{1},k_{2},k_{3})\,.
\end{equation}
The resulting bispectrum is achieved by considering that the
additional shape functions can be defined by using other shape
functions (mentioned in (Renaux-Petel 2012, De Felice and Tsujikawa 2013)) in the Horndeski's
theories, which can be written as
\begin{eqnarray}
\label{eq59}{\cal{A}}_{\Psi}(k_{1},k_{2},k_{3})=\Biggr\{\frac{3}{2}
\Bigg(\frac{1}{c^{2}_{s}}-1-\frac{2\Lambda}{\Sigma}+\frac{6}{\epsilon_{s}}\Big(\frac{\gamma
X\dot{\varphi}}{m^{2}_{pl}HF}\Big)\nonumber\\
-\frac{6}{c^{2}_{s}}\Big(\frac{\gamma
X\dot{\varphi}}{m^{2}_{pl}HF\epsilon_{s}}\Big)\Bigg)
\frac{\Big(\prod^{3}_{i=1}k^{2}_{i}\Big)}{K^{3}}+\Bigg(\frac{3}{4}\Big(\frac{1}{c^{2}_{s}}-1\Big)\Bigg) \nonumber\\
\bigg(\frac{2}{K}\sum_{i>j}k^{2}_{i}k^{2}_{j}-\frac{1}{K}
\sum_{i\neq j}k^{2}_{i}k^{3}_{j}\bigg)+\frac{1}{8}\Big(\frac{1}{c^{2}_{s}}-1\Big)\nonumber\\
\bigg(\sum_{i}k^{3}_{i}+\frac{4}{K}
\sum_{i>j}k^{2}_{i}k^{2}_{j}-\frac{2}{K^{2}}\sum_{i\neq
j}k^{2}_{i}k^{3}_{j}\bigg)\Biggr\}\,.
\end{eqnarray}
The follwing parameter gives the amplitude of the non-Gaussianity
\begin{equation}
\label{eq60}f_{NL}=\frac{10}{3}\frac{{\cal{A}}_{\Psi}}{\Sigma^{3}_{i=1}k^{3}_{i}}\,.
\end{equation}
Following Refs. (Renaux-Petel 2012, De Felice and Tsujikawa 2013) we
introduce the following shapes
\begin{eqnarray}
\label{eq61}{\cal{S}}^{equil}_{*}=\frac{18}{13}\Biggl[3\bigg(\frac{2}{K}\sum_{i>j}k^{2}_{i}k^{2}_{j}-\frac{1}{K}
\sum_{i\neq j}k^{2}_{i}k^{3}_{j}\bigg)-\nonumber\\
\bigg(\sum_{i}k^{3}_{i}+\frac{4}{K}
\sum_{i>j}k^{2}_{i}k^{2}_{j}-\frac{2}{K^{2}}\sum_{i\neq
j}k^{2}_{i}k^{3}_{j}\bigg)\Biggl]\nonumber\\
-\frac{216}{13}\Bigg(\frac{\prod^{3}_{i=1}k^{2}_{i}}{K^{3}}\Bigg)\,,
\end{eqnarray}
and
\begin{eqnarray}
\label{eq62}{\cal{S}}^{ortho}_{*}=\hspace{6cm}\nonumber\\
\frac{12}{14-13\beta}\Biggr[\bigg(3-\frac{9}{2}\beta\bigg)
\bigg(\frac{2}{K}\sum_{i>j}k^{2}_{i}k^{2}_{j}-\frac{1}{K}
\sum_{i\neq
j}k^{2}_{i}k^{3}_{j}\bigg)\nonumber\\+\Big(\frac{3}{2}\beta-1\Big)\bigg(\sum_{i}k^{3}_{i}+\frac{4}{K}
\sum_{i>j}k^{2}_{i}k^{2}_{j}-\frac{2}{K^{2}}\sum_{i\neq
j}k^{2}_{i}k^{3}_{j}\bigg)\nonumber\\
+18\beta\frac{\Big(\prod^{3}_{i=1}k^{2}_{i}\Big)}{K^{3}}\Biggr]\,,
\end{eqnarray}
which are orthogonal. Now, we rewrite equation (59) in terms of
${\cal{S}}^{equil}_{*}$ and ${\cal{S}}^{ortho}_{*}$ (Renaux-Petel 2012) as
\begin{equation}
\label{eq63}{\cal{A}}_{\Psi}=a_{1}{\cal{S}}^{equil}_{*}+a_{2}{\cal{S}}^{ortho}_{*}\,,
\end{equation}
where
\begin{eqnarray}
\label{eq64}a_{1}=\frac{13}{12}\Bigg[\frac{1}{24}\bigg(1-\frac{1}{c^{2}_{s}}\bigg)(2+3\beta)
+\frac{\Lambda}{12\Sigma}(2-3\beta)-\nonumber\\
\frac{1}{6\epsilon_{s}}\bigg(\frac{\gamma
X\dot{\varphi}}{m^{2}_{pl}HF}\bigg)(2-3\beta)
+\frac{1}{3\epsilon_{s}c^{2}_{s}}\bigg(\frac{\gamma
X\dot{\varphi}}{m^{2}_{pl}HF}\bigg)\Bigg]\,,
\end{eqnarray}
and
\begin{equation}
\label{eq65}a_{2}=\frac{14-13\beta}{12}\Bigg[\frac{1}{8}\bigg(1-\frac{1}{c^{2}_{s}}\bigg)
-\frac{\Lambda}{4\Sigma}+\frac{1}{2\epsilon_{s}}\bigg(\frac{\gamma
X\dot{\varphi}}{m^{2}_{pl}HF}\bigg)\Bigg]\,.
\end{equation}

We can obtain the amplitudes of the non-Gaussianity in the
equilateral and orthogonal configurations from equations (60)-(62)
as follows
\begin{eqnarray}
\label{eq66}f^{equil}_{NL}=\Bigg(\frac{130}{36\sum^{3}_{i=1}k^{3}_{i}}\Bigg)
\Bigg[\frac{1}{24}\bigg(1-\frac{1}{c^{2}_{s}}\bigg)(2+3\beta)
\nonumber\\+\frac{\Lambda}{12\Sigma}(2-3\beta)-\frac{1}{6\epsilon_{s}}\bigg(\frac{\gamma
X\dot{\varphi}}{m^{2}_{pl}HF}\bigg)(2-3\beta)
\nonumber\\+\frac{1}{3\epsilon_{s}c^{2}_{s}}\bigg(\frac{\gamma
X\dot{\varphi}}{m^{2}_{pl}HF}\bigg)\Bigg] {\cal{S}}^{equil}_{*}\,,
\end{eqnarray}
\begin{eqnarray}
\label{eq67}f^{ortho}_{NL}=\Bigg(\frac{140-130\beta}{36\sum^{3}_{i=1}k^{3}_{i}}\Bigg)
\Bigg[\frac{1}{8}\bigg(1-\frac{1}{c^{2}_{s}}\bigg)
-\frac{\Lambda}{4\Sigma}+\nonumber\\
\frac{1}{2\epsilon_{s}}\bigg(\frac{\gamma
X\dot{\varphi}}{m^{2}_{pl}HF}\bigg)\Bigg] {\cal{S}}^{ortho}_{*}\,.
\end{eqnarray}
Considering that at $k_{1}=k_{2}=k_{3}$ limit, both the equilateral
and orthogonal configurations have a maximal signal, we obtain the
non-linear parameters in this limit. The results are as
\begin{eqnarray}
\label{eq68}f^{equil}_{NL}=\frac{325}{18}\Bigg[\frac{1}{24}\bigg(1-\frac{1}{c^{2}_{s}}\bigg)(2+3\beta)
+\frac{\Lambda}{12\Sigma}(2-3\beta)\nonumber\\-\frac{1}{6\epsilon_{s}}\bigg(\frac{\gamma
X\dot{\varphi}}{m^{2}_{pl}HF}\bigg)(2-3\beta)
+\frac{1}{3\epsilon_{s}c^{2}_{s}}\bigg(\frac{\gamma
X\dot{\varphi}}{m^{2}_{pl}HF}\bigg)\Bigg]\,,
\end{eqnarray}
and
\begin{eqnarray}
\label{eq69}f^{ortho}_{NL}=\frac{10}{9}\Big(\frac{65}{4}\beta+\frac{7}{6}\Big)
\Bigg[\frac{1}{8}\bigg(1-\frac{1}{c^{2}_{s}}\bigg)
-\frac{\Lambda}{4\Sigma}+\nonumber\\
\frac{1}{2\epsilon_{s}}\bigg(\frac{\gamma
X\dot{\varphi}}{m^{2}_{pl}HF}\bigg)\Bigg]\,.
\end{eqnarray}
After calculation of perturbations and possible non-Gaussianity of
these perturbations we compare our results with observations in the
next section.

\section{Confrontation with Observational Data}

In this section we perform a numerical analysis on the parameter
space of our generalized G-inflation model and compare the results
with Planck2015 observational data. To this end, we adopt a
potential as $V=\frac{\lambda}{4}\varphi^{4}$ and  we set
${\cal{K}}=1$ and $\gamma(\varphi)=\varphi/M^{4}$. To perform the
numerical analysis we assume $M\simeq10^{-5}m_{pl}$,
$\mu\simeq3\times10^{-8}m_{pl}$ and $N=60$. Now, by solving the
integral of equation (30), we obtain the value of the Higgs field at
the horizon crossing of the physical scales. After that, by using
this obtained value we can find the scalar spectral index,
tensor-to-scalar ratio and the amplitudes of the equilateral and the
orthogonal configurations of the non-Gaussianity in terms of $N$,
$\lambda$ and $\xi$. Then, we analyze the model parameter space
numerically. The results are shown in figures.

Figure 1 shows the ranges of the self-coupling parameter of the
Higgs field, $\lambda$, and the non-minimal coupling parameter,
$\xi$, that lead to the observationally viable values of the scalar
spectral index and tensor-to-scalar ratio. In plotting the figures
we have focused on $\lambda<10^{-6}$ and $\xi<2\times 10^{2}$.
Figure shows that, as $\xi$ increases the smaller values of
$\lambda$ are observationally viable. In figure 2 we have plotted
the tensor-to-scalar ratio versus the scalar spectral index in the
background of Planck2015 TT, TE, EE+lowP data. To plot this figure,
we have adopted three sample values of the non-minimal coupling
parameter as $\xi=50$, $\xi=80$ and $\xi=100$. Our numerical
analysis shows that this generalized G-inflation model is consistent
with Planck2015 data if $10^{-7}\leq\lambda\leq2\times10^{-6}$ for
$\xi=50$, $10^{-7}\leq\lambda\leq5\times10^{-6}$ for $\xi=80$ and
$10^{-7}\leq\lambda\leq7\times10^{-6}$ for $\xi=100$. Note that the
presence of the Galileon-like interaction and the NMC effect in this
model cause a reduction of the tensor-to-scalar ratio in comparison
to the standard situation. We have also studied the amplitudes of
the non-Gaussianity in both the equilateral and orthogonal
configurations numerically. The results are shown in figures 3 and
4. We have analyzed $f_{NL}^{equil}$ and $f_{NL}^{ortho}$ in the
ranges of the parameters used in studying $r$ and $n_{s}$. Figures 3
and 4 show that in the ranges $\lambda<10^{-6}$ and $\xi<2\times
10^{2}$, both equilateral and orthogonal non-Gaussianities are
consistent with Planck2015 TTT, EEE, TTE and EET data. As these
figures show, in this generalized G-inflation model, it is possible
to have large non-Gaussianity in some subspaces of the model
parameter space. From our analysis we can say that if
we consider a generalized G-inflation model, depending on the values
of $\xi$, it is possible to have $\lambda<10^{-6}$ (specially,
$\lambda\sim 10^{-13}$ which is well in the range of CMB
result~(Liddle and Lyth 2000)). This means that, if we adopt smaller
values of $\xi$, it is possible to reduce the self-coupling of the
Higgs sector in order to reach the energy scale of inflation in this setup.
This is an important results since it provides a possible mechanism
for reduction of the Higgs self-coupling as an inflaton.

\begin{figure*}
\begin{center}
\includegraphics{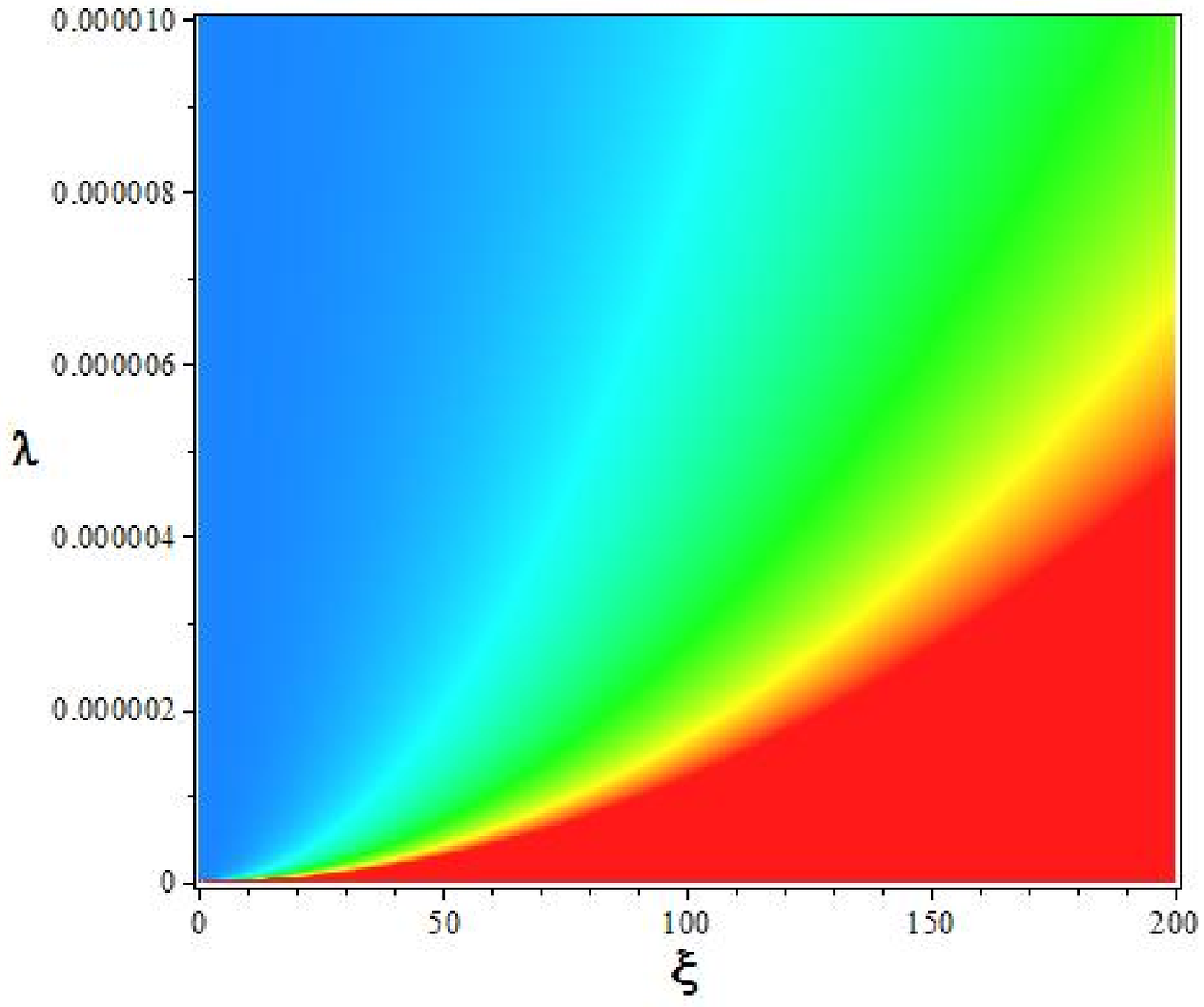} \hspace{0.0cm} \includegraphics{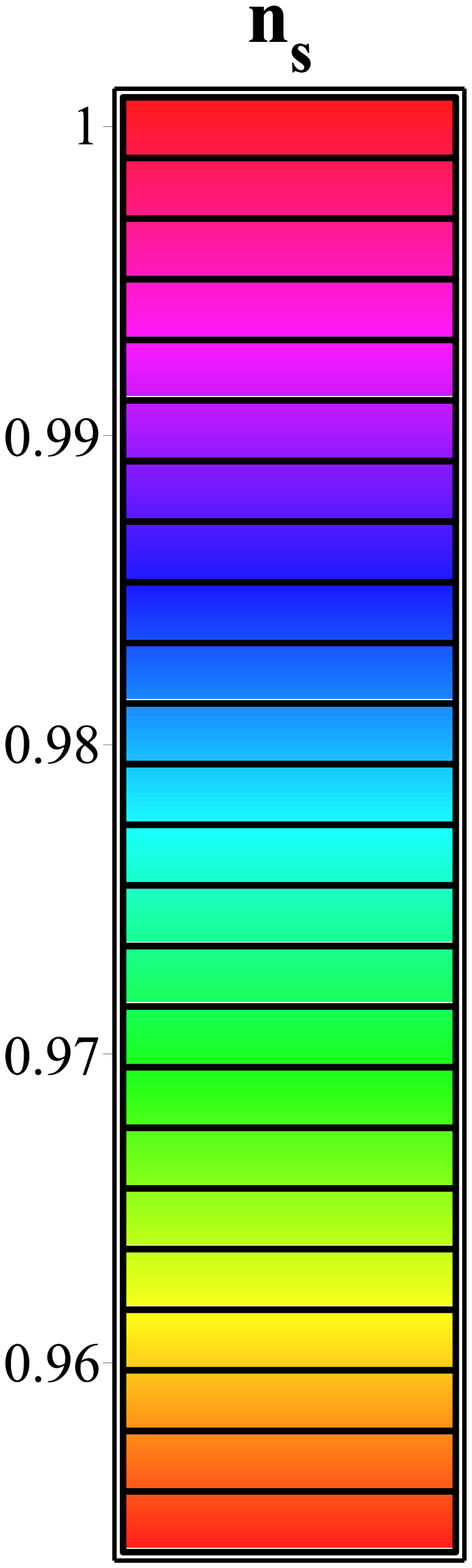}
\hspace{5cm}\hspace{0.0cm} \includegraphics{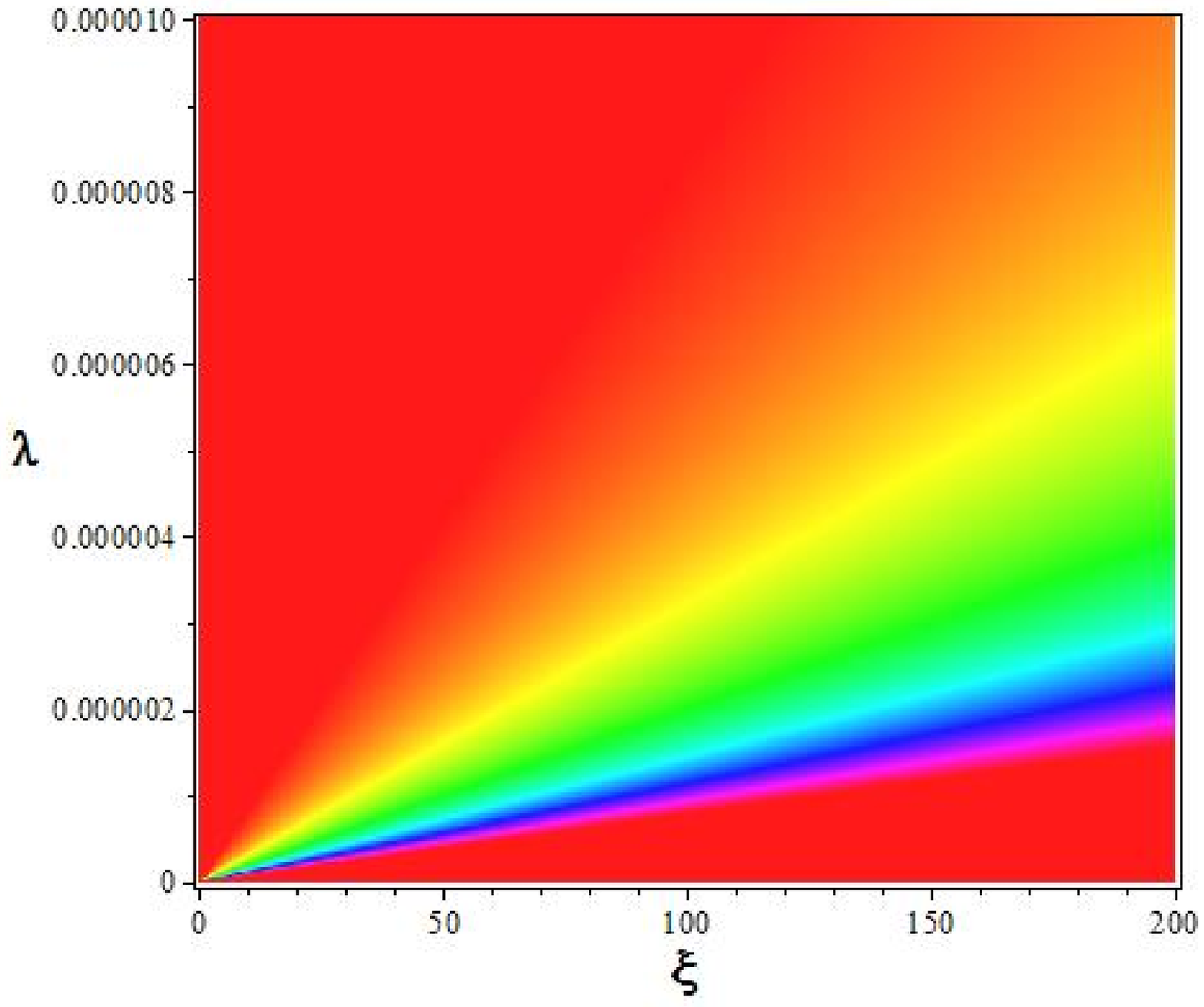}\vspace{7cm}
\includegraphics{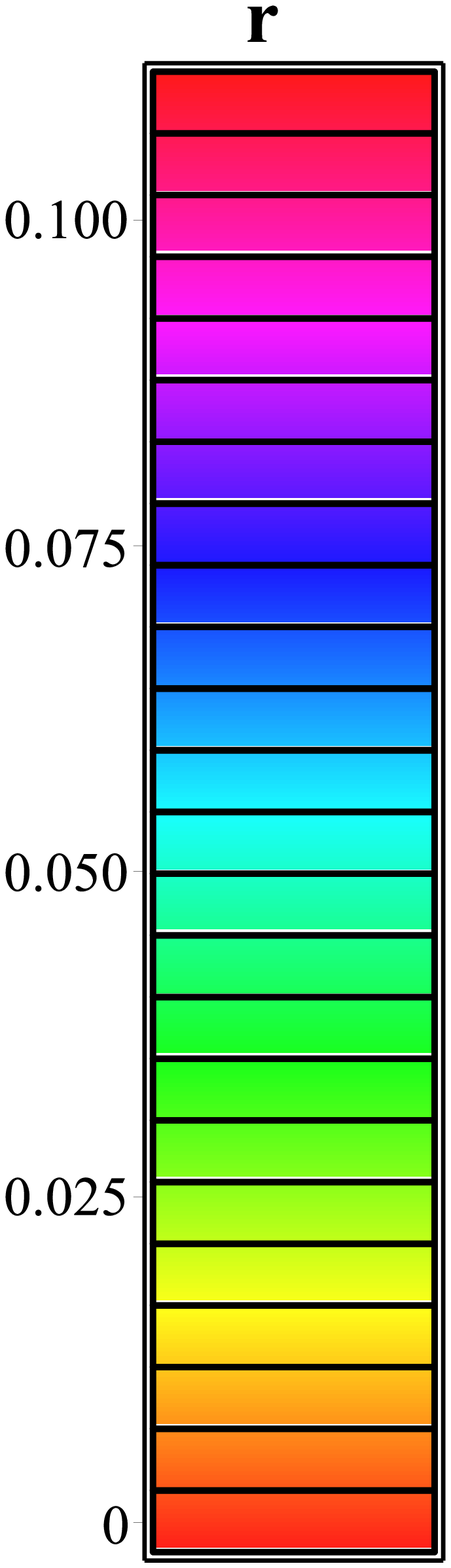} \hspace{5cm} \caption{\small {Ranges of $\lambda$ and
$\xi$ leading to the observationally viable values of the scalar
spectral index (left panel) and the tensor-to-scalar ratio (right
panel) for a generalized Higgs Galileon model. We note that
consistency with observations in this generalized model requires
enhancement of $\lambda$.}}
\end{center}
\end{figure*}

\begin{figure*}
\begin{center}\includegraphics{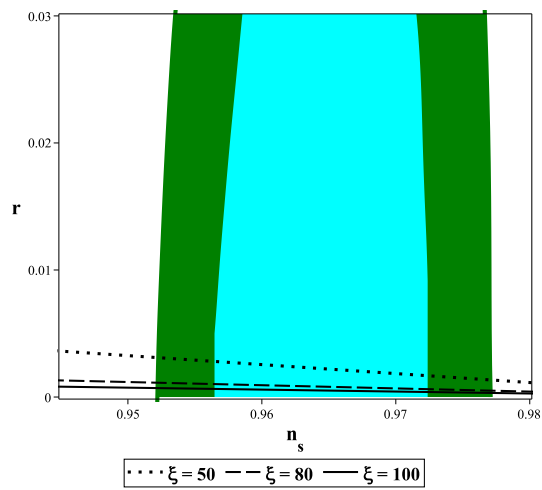} \vspace{7cm}
\end{center}
\caption{\small {Tensor-to-scalar ratio versus the scalar spectral
index for a generalized Higgs model, in the background of Planck2015
TT, TE, EE+lowP data. The figure is plotted with N=60.}}
\end{figure*}

\begin{figure*}
\begin{center}
\includegraphics{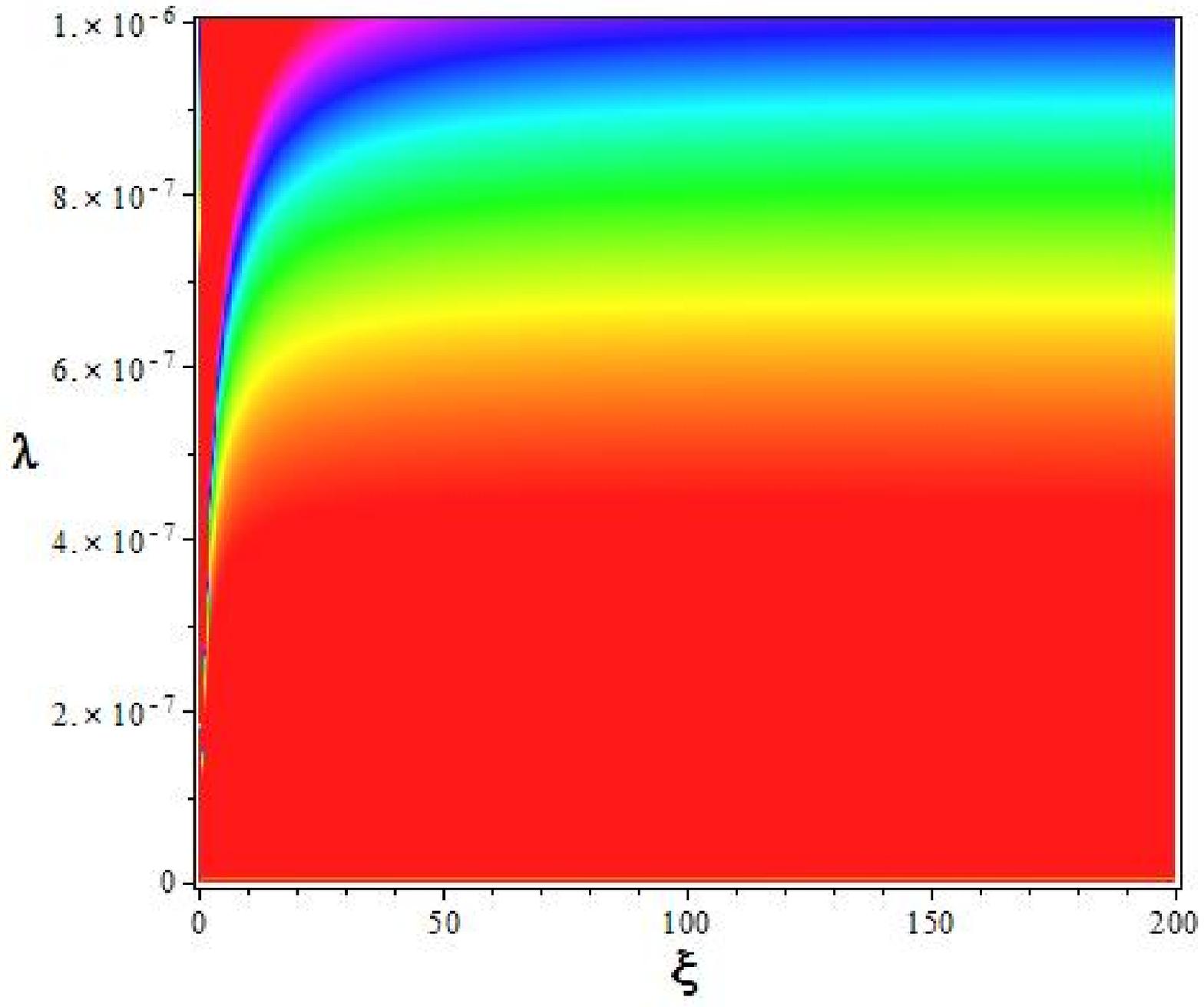} \hspace{0.0cm} \includegraphics{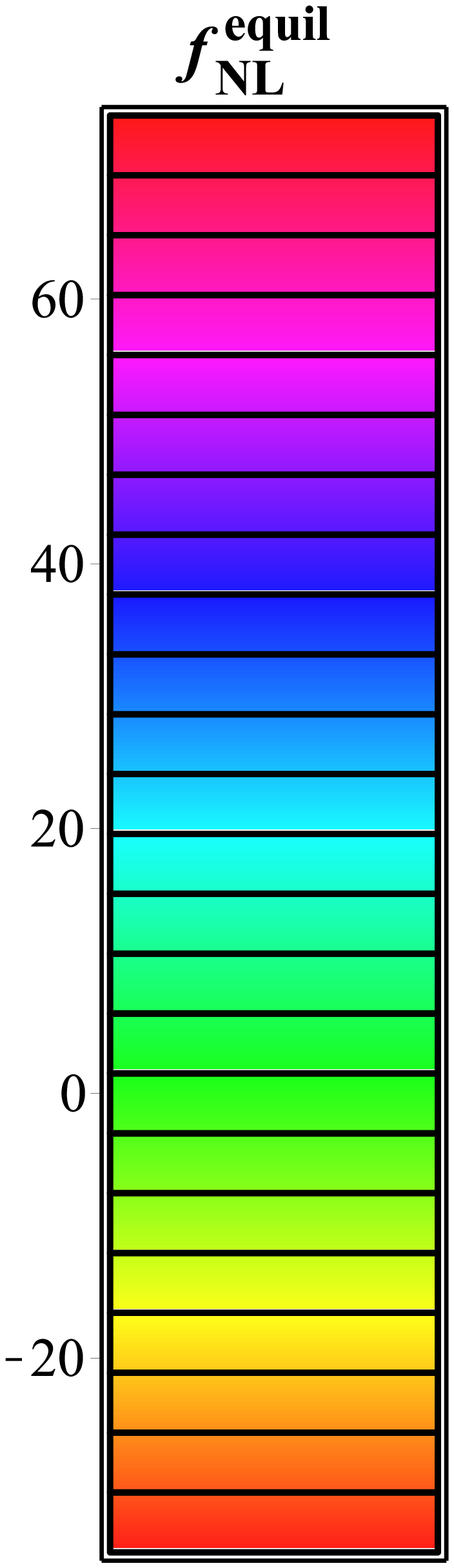}
\hspace{5cm}\hspace{0.0cm} \includegraphics{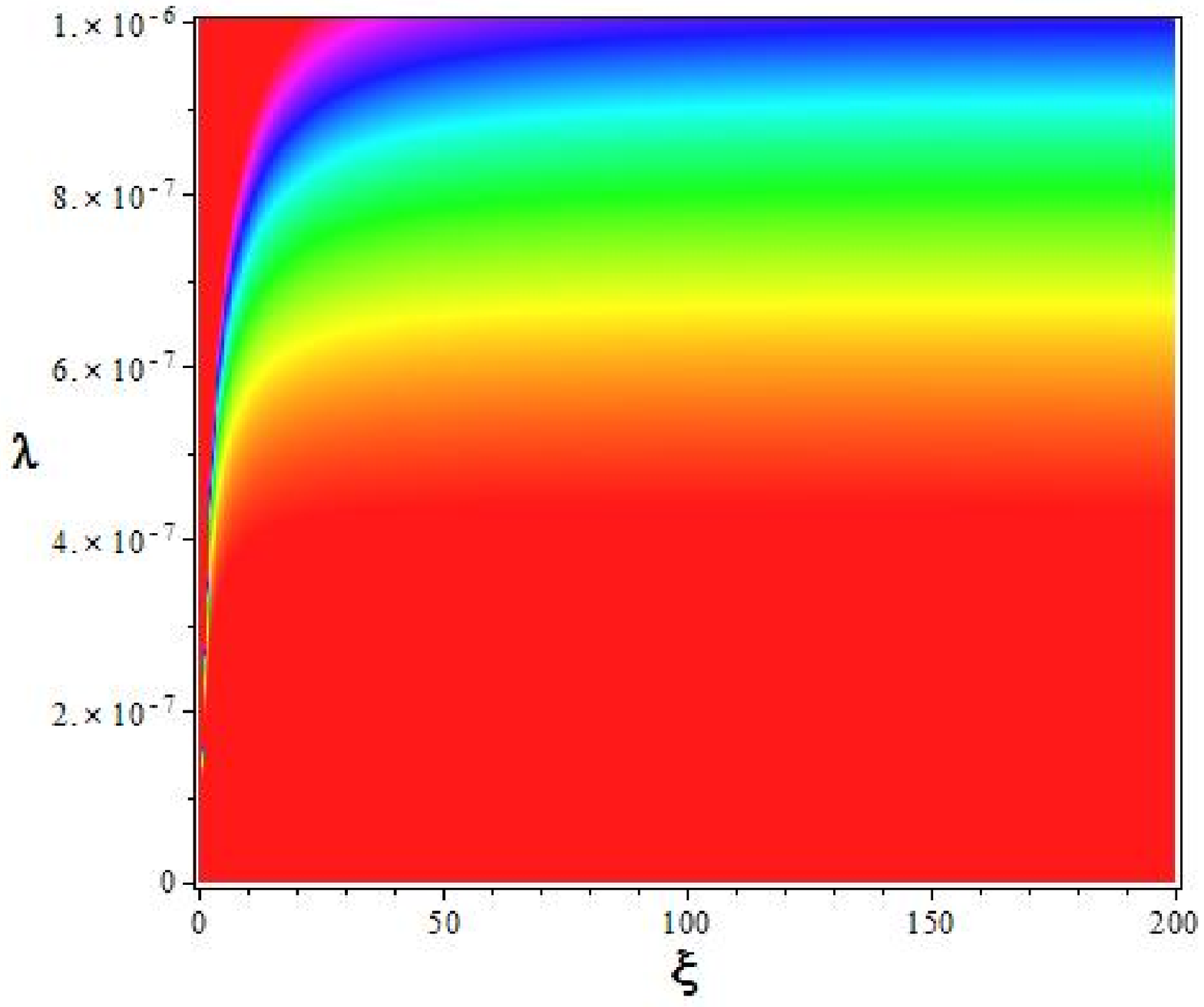}\vspace{7cm}
\includegraphics{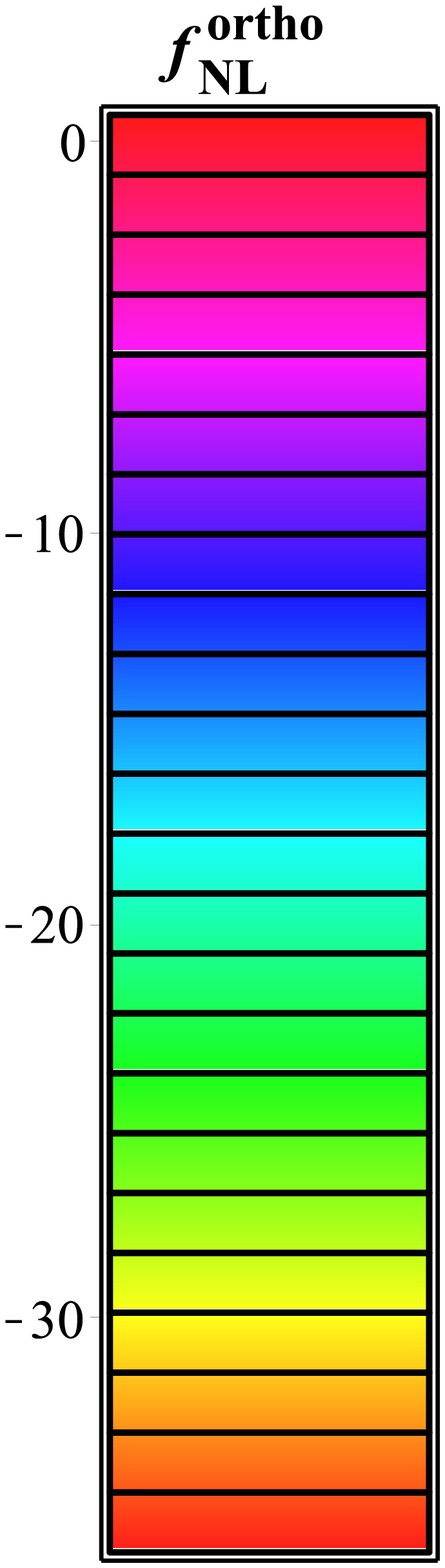} \hspace{5cm} \caption{\small {Ranges of
$\lambda$ and $\xi$ leading to observationally viable values of the
amplitudes of the equilateral (left panel) and orthogonal (right
panel) configurations of the non-Gaussianity for a generalized Higgs
inflation model. In both panels all the adopted ranges are
consistent with Planck 2015 observational data. }}
\end{center}
\end{figure*}

\begin{figure*}
\begin{center}\includegraphics{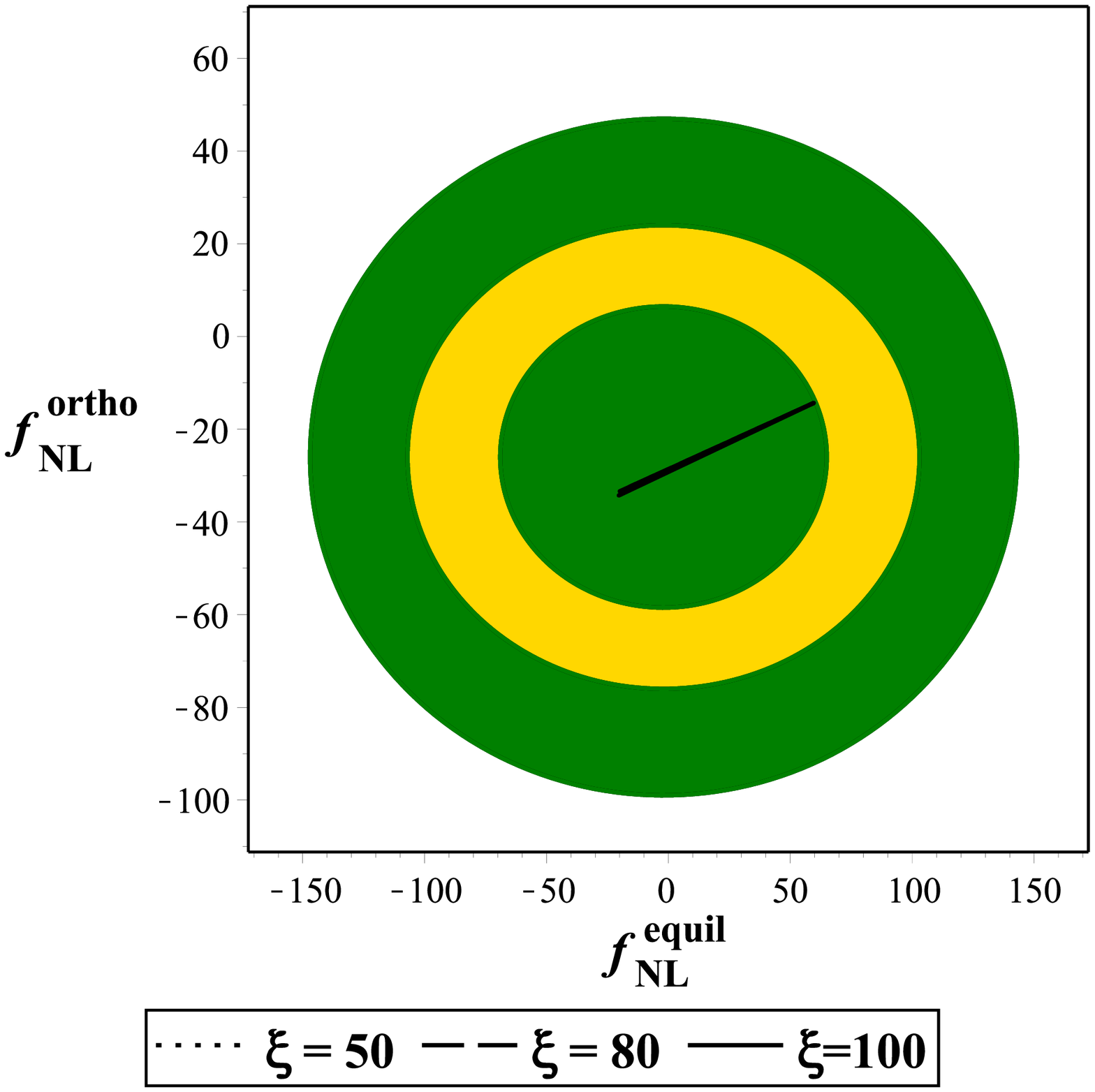} \vspace{7cm}
\end{center}
\caption{\small {Amplitude of the orthogonal configuration of the
non-Gaussianity versus the amplitude of the equilateral
configuration for a generalized Higgs inflation in the background of
Planck2015 TTT, EEE, TTE and EET data.  The figure is plotted with
N=60. Note that the diagrams for all three values of the non-minimal
coupling are too close to be distinguished from each other in this
figure. }}
\end{figure*}

\section{Summary}

In this paper we have studied the cosmological inflation in a
generalized G-inflation model. We have studied the effects of the
Galileon interaction and the non-minimal coupling on the energy
scale of the Higgs inflation. In this regard we have adopted the
non-minimal coupling function as $\xi\varphi^{2}$ and other
functions as $V=\frac{\lambda}{4}\varphi^{4}$, ${\cal{K}}=1$ and
$\gamma(\varphi)=\frac{\varphi}{M^{4}}$. We have obtained the
background dynamics and then we have treated the perturbations in
this generalized setup in details. By expanding the action up to the
second order, we have obtained the scalar and tensor spectral
indices and tensor-to-scalar ratio in this generalized G-inflation
model. In this respect, we have shown that the presence of the
Galileon effect modifies the consistency relation. By calculating
the cubic action and the three point correlation function, we have
studied the non-Gaussian feature of perturbations in this setup. We
have also obtained the non-linear parameters in both equilateral and
orthogonal configurations of the non-Gaussianity at
$k_{1}=k_{2}=k_{3}$ limit. Finally, we have performed a numerical
analysis on the model's parameter space to obtain some constraints
on the parameters. We have studied $n_{s}$, $r$, $f_{NL}^{equil}$
and $f_{NL}^{ortho}$ numerically. Our numerical analysis shows that
if we consider the non-minimal coupling and Galileon-like
interactions, it is possible to control the values of the
self-coupling parameter $\lambda$. Actually, in this extended model, depending on the
values of the non-minimal coupling, we were able to reduce the values of
$\lambda$ from interval $0.11<\lambda<0.27$ to $\lambda<10^{-6}$. In
fact, if we adopt smaller values of $\xi$, it is possible to reduce
the energy scale (self-coupling) of the Higgs sector in order to reach the energy scale of
inflation ($\lambda\sim 10^{-13}$) in this setup. Therefore, by
reducing the order of $\lambda$ and approaching the energy scale of
the inflation era, the Higgs field can be considered to be an
inflaton.

\acknowledgments

The work of K. Nozari has been supported financially by Research
Institute for Astronomy and Astrophysics of Maragha (RIAAM) under
research project number 1/4717-71.\\

\end{document}